\documentstyle{mn}  

\begin{document}  
 
\input epsf  
\input psfig

\def\spose#1{\hbox to 0pt{#1\hss}}
\def\simlt{\mathrel{\spose{\lower 3pt\hbox{$\mathchar"218$}}
     \raise 2.0pt\hbox{$\mathchar"13C$}}}
\def\simgt{\mathrel{\spose{\lower 3pt\hbox{$\mathchar"218$}}
     \raise 2.0pt\hbox{$\mathchar"13E$}}}
\def\deg{^\circ}

\title{The Local Environment of HII Galaxies} 

\author[Telles \& Maddox] {Eduardo Telles$^{1}$ 
\and Steve Maddox $^{2,3}$ \\ 
1.		Observat\'orio Nacional,
		Rua Jos\'e Cristino, 77,
		20921-400 - Rio de Janeiro -
		BRASIL \\
2. Institute of Astronomy, Madingley Road, Cambridge  
CB3 0HA, UK. \\
3. School of Physics and Astronomy, University of Nottingham,
Nottingham, NG7 2RD, UK. \\
etelles@on.br, sjm@ast.cam.ac.uk}

\maketitle  
\def\Mpc{\thinspace\hbox{Mpc}}
\def\kmsec{\thinspace\hbox{$\hbox{km}\thinspace\hbox{s}^{-1}$}}
\def\kmsecmeg{\thinspace\kmsec\Mpc$^{-1}$}
\def\msun{\thinspace\hbox{$\hbox{M}_{\odot}$}}
\def\reference#1#2#3#4#5{\tenp\par\noindent\hangindent 3em
              #1, #2. {\tenpsl #3\/}, {\tenpb #4,}
\thinspace\hbox{#5}}
\def\etal   {et\nobreak\ al.\ }
\def\aj     {Astron.\nobreak\ J.\nobreak\ }
\def\mn     {Mon.\ Not.\ R.\ astr.\nobreak\ Soc.\nobreak\ }
\def\apjs   {Astro\-phys.\nobreak\ J.\ Suppl.\nobreak\ }
\def\apj    {Astro\-phys.\nobreak\ J.\nobreak\ }
\def\aanda  {Astr.\ Astro\-phys.\nobreak\ }
\def\apjl   {Astro\-phys.\nobreak\ J.\ Lett.\nobreak\ }
\def\aandas {Astr.\ Astro\-phys.\nobreak\ Suppl.\nobreak\ }
\def\rmx    {Rev.\ Mex.\ Astr.\ Astrofis.\nobreak\ }

\begin{abstract} 

We have carried out an investigation of the environments of low
redshift HII galaxies by cross-correlating their positions on the sky
with those of faint field galaxies in the Automatic Plate Measuring
Machine catalogues.  
We address the question of whether violent star formation in HII
galaxies is induced by low mass companions by statistically estimating
the mean space density of galaxies around them.
We argue that even if low mass companions were mainly intergalactic HI
clouds, their optical counterparts should be detectable at faint
limits of the APM scans.

A significantly positive signal is detected for the HII galaxy-APM
galaxy angular cross-correlation function, but the amplitude is poorly
determined. 
The projected cross-correlation function has higher signal-to-noise,
and suggests that the amplitude is slightly lower than for normal field
galaxies. 
This implies that these bursting dwarf galaxies inhabit slightly lower
density environments to those of normal faint field galaxies,
consistent with other studies of emission line galaxies.
This suggests that in these dwarf starburst galaxies, star formation
is not triggered by unusually strong tidal interactions, and may have
a different origin.

\end{abstract}


\section{Introduction}

\begin{center}
\begin{figure*}
\hspace*{5mm}\psfig{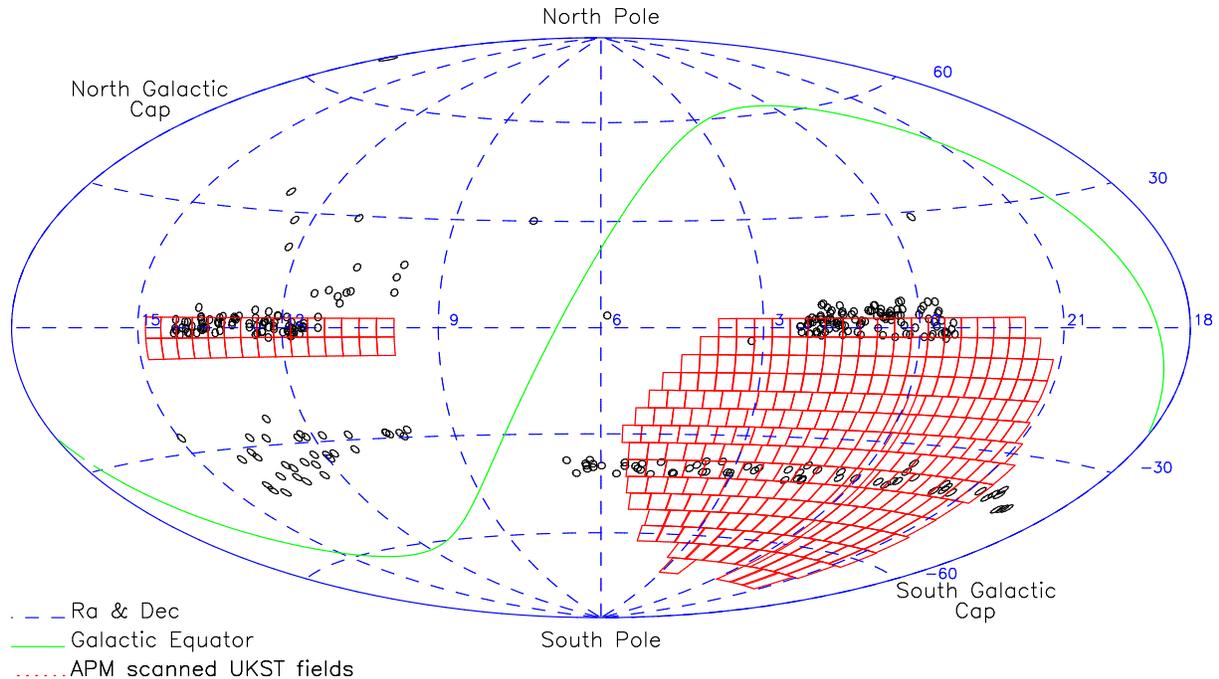} 
\caption{\label{skyplot}
The distribution of HII galaxies and APM data on the sky in an equal area
projection in equatorial coordinates.  The solid circles are the
positions of the HII galaxies.  The square fields are the APM scans of
the UK schmidt plates.  The dashed lines show lines of constant RA and
DEC.} 

\end{figure*}
\end{center}

The starburst phenomenon is observed in a large number of
extragalactic objects from giant HII regions in irregular galaxies and
late type spirals, to galaxies entirely dominated by the violent
massive star formation region as in the case of Starburst Galaxies.
The class of starburst galaxies comprises a large range in luminosity, 
mass, heavy element and dust content, as well as morphology.
Classical starburst or nuclear starburst galaxies typically have an
intense region of violent star formation in the nucleus of an
otherwise normal spiral galaxy (Balzano 1983).  At the high luminosity
end of starburst galaxies, ultraluminous IRAS galaxies (Soifer et
al. 1987) are strongly interacting giant systems (Melnick \& Mirabel
1990) where most of the radiation is emitted in the far-infrared due
to reprocessed UV radiation by the large content of dust particles.
HII galaxies, on the other hand, are dwarf galaxies in a bursting
phase of star formation of low luminosity and mass, low heavy element
abundance and low dust content where the triggering mechanism of the
present episode of violent star formation is not as obvious (Telles \&
Terlevich 1995).

Earlier searches for bright galaxies near to HII galaxies found a
deficit of $L>L^*$ galaxies within 1 Mpc (Campos-Aguilar \& Moles
1991, Campos-Aguilar \etal 1993, Vilchez 1995, Pustil\'nik \etal 1995;
Telles \& Terlevich,1995).
HII galaxies are not associated with giant galaxies, therefore they 
are not tidal debris of strongly interacting systems.
Auto-correlation analyses of strong emission line galaxies (Iovino,
Melnick \& Shaver 1988, Rosenberg \etal 1994, Loveday \etal 1999) find
a low clustering amplitude, suggesting that HII galaxies tend to
populate regions of low galactic density.

Their optical properties are dominated by the massive star forming
region, as shown by their strong emission line spectra superposed on a
weak blue continuum.
The properties of the underlying galaxies in these systems are similar
to late type dwarf galaxies such as dwarf irregulars or low surface
brightness dwarfs (Telles \& Terlevich 1997).
The most luminous HII galaxies, classified as Type I's by Telles,
Melnick and Terlevich (1997), show signs of disturbed morphology such
as distorted outer isophotes, tails or irregular fuzz, while the low
luminosity Type II's are regular and compact.
Although there is a clear case for a morphology-luminosity relation,
neither type of HII galaxies shows conspicuous evidence of bright
companions in their neighbourhood.
The few HII galaxies found to have a bright neighbour (maybe by
chance) are all of Type II's of regular morphology, contrary to what
one would expect if interactions caused the morphological disturbances
as seen in Type I's (Telles \& Terlevich 1995).

A popular hypothesis is that interactions between galaxies are the
triggers of starbursts and they may also cause the current burst of
star formation in HII galaxies.  Giant galaxies, however, are not
found in the immediate vicinity of HII galaxies, thus are improbable
candidates for triggering agents.  A possible and appealing
alternative was presented by Melnick (1987) who proposed that high
resolution 21cm maps were needed to investigate the role of collisions
between intergalactic neutral hydrogen clouds in the formation of
these objects.  Brinks (1990) also hypothesized that other dwarfs or
intergalactic HI clouds could be the triggering agents.  Taylor \etal
(1995, 1996) using the VLA detected 12 HI companions around 21 HII
galaxies, while only 4 HI-rich companions were detected around a
control sample of 17 quiescent low surface brightness dwarfs (Taylor
1997).  As also pointed out by these authors, some questions remain
intriguing from this: Why are these 9 out of the 21 HII galaxies with
{\em no} companions violent forming stars now ('bursting')?  Why are
these 4 out of the 17 LSBGs with companions {\em not} 'bursting'?

HI surveys find that all the HI detections have an optical
counterpart.  That is, all the sources found in 21 cm surveys are
nothing else than normal galaxies (c.f. Zwaan \etal 1997, Zwaan, 1999,
Hoffman, 1999).  No free floating intergalactic HI clouds were
detected in such surveys.  Thus, we have carried out a further
investigation of the galaxy environments of a unbiased sample of over
160 low redshift HII galaxies by cross-correlating their accurate
position in the sky to faint field galaxies ($15 < b_{\rm J} < 20.5$)
in the Automatic Plate Measuring Machine (APM) galaxy catalogue.  
For the mean redshift of our HII galaxy sample we 
detect galaxies down to ${\rm M_B} \approx -14.5$.  
Using the relation between optical magnitude M$_B$ and HI mass
${\cal M}_{\rm HI}$ for late type galaxies, given by Rao \& Briggs (1993)
[$\log{\cal M}_{\rm HI}{\rm \msun} = 2.72 - 0.36 {\rm M_B}$], we estimate
that we are not missing any cloud with mass greater than 10$^8$\msun. 
This is comparable to  the lower limits of HI companions found by
Taylor \etal (1995, 1996).
Hence our present study should detect any possible low mass candidates
to act as tidal triggers.

In Section~\ref{data} we describe in more detail the data-sets used in
the present work and we present the details of our calculations in
Section~\ref{analysis}.  Finally, we show some of our conclusions in 
Section~\ref{conclusion}.

\section{Data}\label{data}

\subsection{HII Galaxy Sample} 

The HII galaxy sample used in this paper is taken from the {\em
Spectrophotometric Catalogue of HII Galaxies} (SCHG; Terlevich \etal
1991).
Most of the objects in this database have been selected from the
Tololo survey (Smith, Aguirre \& Zemelman 1976), and the University of
Michigan survey (MacAlpine \& Williams 1981).
The catalogue also contains a number of objects which are not
classified as Seyfert galaxies selected from the Markarian list of
galaxies with strong ultraviolet continuum (Markarian, Lipovetskii \&
Stepanyan 1981 and references therein), as well as some blue objects
of Zwicky's catalogue of compact galaxies (Zwicky 1971).
The total catalogue contains spectra of over 400 emission line objects
found in objective prism surveys using IIIa-J emulsion through their
[OIII]$\lambda\lambda$4959,5007 and/or [OII]$\lambda\lambda$3726,3729
lines.
From these, about 300 are HII galaxies.  
The remainder are Giant HII regions, and Starburst nuclei or emission
line objects classified as Seyfert nuclei from their position in the
emission-line-ratio diagnostic diagrams.

Most of the objects in this sample cover only two specific regions of
the sky.  
For instance, the Michigan survey covers a 10$\deg$ band around the
celestial equator, while the Tololo survey concentrates in the region
$-27\deg < \delta < -43 \deg$.
This is illustrated in Figure~\ref{skyplot} which shows the
distribution of our HII galaxies in the sky.
For the present study we ended up with 163 HII galaxy centres for
which there are APM scanned UK Schmidt plates.
The actual centres for the HII galaxy were carefully identified for
each APM field, thus assuring that the HII galaxy is not counted as a
companion of itself. 

\begin{figure}
\protect\centerline{
\epsfxsize=3.5in\epsffile{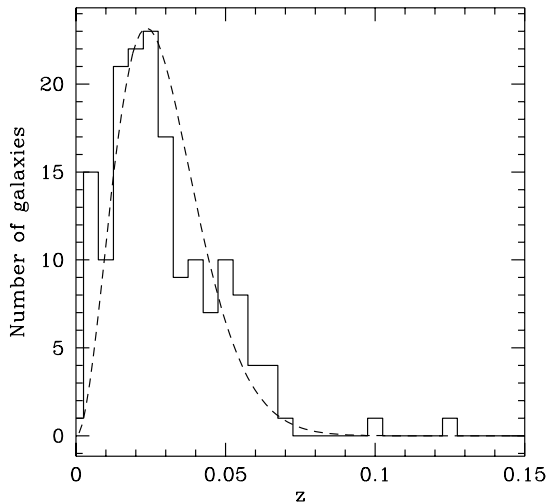}
}
\caption{
The distribution of redshifts for the HII galaxies (histogram) and the model
distribution (equation~\ref{eq:nz} 
used to calculate the model correlation function (dashed line). 
}
\label{nz}
\end{figure}

The redshift distribution of these galaxies is plotted in
Figure~\ref{nz}, which also shows the best-fit model redshift
distribution of the form 
\begin{equation} 
 N(z) \propto { z^2  \over z_c^3} 
{\rm exp} \left [ - \left( {z \over z_c}
\right)^{3/2} \right ] 
\label{eq:nz} 
\end{equation}
The mean redshift is 0.03.  The typical absolute magnitude is M$_{\rm
B} \approx -18$. Throughout this paper we use the current value 
of H$_0 = 65$ \kmsecmeg
(Suntzeff \etal 1998).


\subsection{APM galaxy sample}

Our sample of faint field galaxies was selected from the APM Galaxy
Survey, which is described in detail by Maddox et al.\ (1990).
The sky covered by the APM galaxy survey has been extended since the
original publication of the survey: the south galactic pole part of
the survey now covers a solid angle of $6250$ square degrees, and is
based on 269 UKS J plates scanned by the APM machine; the north
galactic cap area covers 750 square degrees from scans of 30 UKS
plates centred with $9^{\rm h} <\alpha < 15^{\rm h} $ and $-5\deg <
\delta < 0\deg$.  
The fields covered are shown by the dotted squares on
Figure~\ref{skyplot}.

The data consist of positions accurate to $\simlt 1''$ and 
magnitudes accurate to $\sim 0.1$ mag for over 50 million
images brighter than a magnitude limit of $b_{\rm J} =21.9$.  
The galaxy photometry has been corrected for several systematic
effects and have no detectable systematic errors more than $\sim 0.04$
magnitudes rms.
The galaxy sample selected from the survey data at a magnitude limit
of $b_{\rm J} =20.5$ has a completeness $\sim 90$--$95\%$, stellar
contamination $\sim 5\%$ and negligible dependence of the galaxy
surface density on declination or galactic latitude (Maddox et al
1996).
At this magnitude limit the redshift distribution is well described by
equation~\ref{eq:nz} with the mean $z=0.16$ (Maddox et al 1996).

We extracted APM measurements for a $2\deg$ square area around 
each of the HII galaxies. 
For the central 10' square we visually cross-checked the APM catalogue
list against images from the Digitized Sky Survey (DSS), and rejected
multiple detections and noise images.
This provides a reliable galaxy catalogue around each HII galaxy.

\section{Analysis}\label{analysis}

\subsection{ The angular cross-correlation function } 

We have measured $w_{hg}(\theta)$, the angular cross-correlation
function between the HII galaxies and an apparent magnitude limited
sample of neighbouring galaxies. 
We estimated $w_{hg}$ by counting the number of galaxies $N_{hg}$ as a
function of angular radius $\theta$ from the central HII galaxy, and
comparing this to the number $N_{hr}$ counted for a catalogue of
uniform random postions. We used ten times as many random points as
galaxies in order to reduce their contribution to the counting
errors, and then rescaled the count down by a factor ten. 
The cross-correlation function is then given by 
\begin{equation} 
w_{hg}(\theta) = {{ N_{hg}(\theta)} \over {N_{hr}(\theta)} }   -1
\label{whg} 
$$\end{equation} 
We also used the simpler direct estimate using the mean surface
density of field galaxies, $\bar {\cal N}$, and the area of each
annulus, 
\begin{equation} 
w_{hg}(\theta) =  {{ N_{hg}(\theta) } \over { n_{cen}~ \bar {\cal N } \pi
((\theta-\Delta)^2 - \theta^2))}}  -1 
\label{whg_d} 
\end{equation}
where $n_{cen}$ is the number of HII galaxies used as centres. 

\begin{figure*}
\protect\centerline{
\hspace*{5mm}\psfig{figure=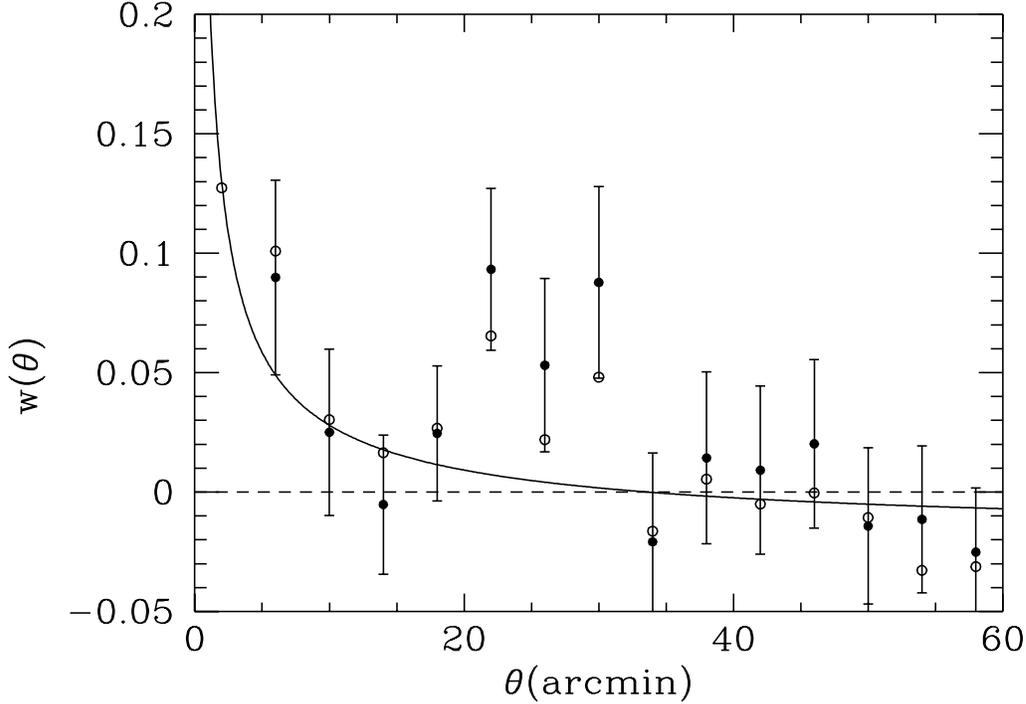,angle=-90,width=16cm} 
}
\vspace{-10mm}
\caption{\label{wtheta}{The angular cross-correlation between
the HII galaxies and the faint APM field galaxies. The filled points
show the results using equation~\ref{whg}, and the open points are
from equation~\ref{whg_d}. The error bars are estimated from 
$\epsilon_w = (1+2\pi\bar {\cal N}
J_2(\theta))/\sqrt{N_{hg}}$. 
The line is the cross-correlation function
predicted assuming that the HII galaxies are a random subsample of 
APM galaxies. }} 
\end{figure*}

The main results are shown in Figure~\ref{wtheta}.  
The filled points show $w$ from equation~\ref{whg} and the open points
from the direct estimator, equation~\ref{whg_d}.
This gave essentially indistinguishable results, showing that there
are no significant systematic biases in our sample.
The error bars are estimated from the Poisson noise in each radial
bin, scaled by the integral of $w$, $\epsilon_w = (1+2\pi\bar {\cal N}
J_2(\theta))/\sqrt{N_{hg}}$, where $J_2 = \int_0^\theta \theta w(\theta)
d\theta $. This is  analogous to the error estimate for the spatial
correlation function $\xi$ suggested by Kaiser (1986), and is a reasonable
approximation for a weakly clustered distribution (see Hamilton 1993
for an extensive analysis of errors in correlation functions).
It can be seen that $w_{hg}$ is significantly positive for angles
$\theta\simlt 30 \arcmin$,  corresponding to an excess of
galaxies near the HII galaxy positions compared to a uniform
distribution. 
This would be expected for any sample of galaxies, since we know that
galaxies are clustered. 

We have calculated the expected cross-correlation function 
assuming that HII galaxies cluster in the same way as normal
galaxies, and this is shown by the  line in Figure~\ref{wtheta}. 
This prediction is based on the APM correlation function for $b_j=20$,
which is well fit by a power law at small scales, $w(\theta) = A
\theta^{1-\gamma}$, with $\gamma = 1.699$ and $ A_{APM} = 0.0284$
(Maddox et al 1996).
We scaled the amplitude by a factor calculated numerically from
Limbers equation (Peebles 1980)
\begin{eqnarray}
w_{hg}(\theta)=\frac{\int_0^{\infty}\int_0^{\infty}N_h(z_h)N_g(z_g)\xi(r_{hg})(\frac{dx}{dz})_h(\frac{dx}{dz})_g\:dz_h
\:dz_g}{\left[\int_0^{\infty}N_h(z)\:dz\right]\left[\int_0^{\infty}N_g(z)\:dz\right]
},
\label{eqn:limber} 
\end {eqnarray}
where $x$ is the comoving coordinate at redshift $z$, $N_h(z_h)$ is the
redshift distribution of the HII galaxy sample, $N_g(z_g)$ is the
redshift distribution of the APM galaxy sample and $\xi(r_{hg})$ is the
spatial correlation function at separation $r_{hg}$. 
The two redshift distributions are given by equation~\ref{eq:nz} with $z_c $
chosen to match the observed distributions for HII galaxies and 
the APM galaxies at
$b_J=20.5$. 
This gave the predicted amplitude $ A_{hg} = 0.01399$.  
We also included a constant offset to correct for the integral constraint
within the 2$\deg$ fields
(Smith, Boyle \&Maddox 1995), giving a final $w(\theta) = 0.01399
(\theta^{-0.699} - 1.5)$, where $\theta$ is in degrees.

Figure~\ref{wtheta} shows that the estimated $w_{hg}(\theta)$ for HII
galaxies is consistent with the assumption that HII galaxies are
clustered in the same way as normal field galaxies.  
The uncertainty in the estimated amplitude is rather large, but we can
rule out the suggestion that HII galaxies are anticorrelated with
field galaxies. 
It is also clear that there is not a large excess of near neighbours around
the HII galaxies compared to normal galaxies. 
This appears incompatible with suggestion that HII galaxies are triggered
by tidal interactions with nearby low-mass galaxies.  


\subsection{ The projected cross-correlation function } 

\begin{figure*}
\protect\centerline{
\hspace*{5mm}\psfig{figure=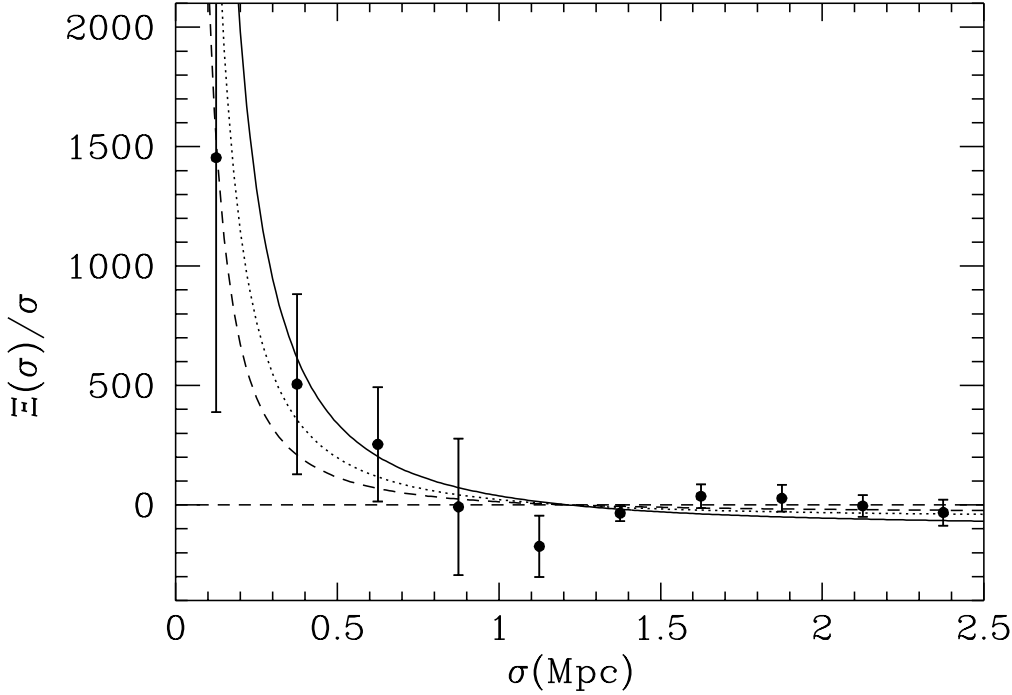,angle=-90,width=16cm} 
}
\vspace{-10mm}
\caption{\label{wsigmar}
The  projected cross-correlation between the HII galaxies and
the faint APM field galaxies, $\Xi_{hg}(\sigma)/\sigma$. The points
show our measurements, with error bars estimated from the scatter
between centres. The solid,
dotted and dashed lines show the predicted $\Xi$ for $r_0=5.1, 3.7$
and $2.7 h^{-1}$ Mpc respectively. 
}
\end{figure*}

Since we know the redshift to each HII galaxy, we can estimate the
correlation function using the projected physical separation,
$\sigma$.
The resulting projected cross-correlation function, $\Xi_{hg}$, is an
integral over the spatial correlation function $\xi_{hg}$,
\begin{equation} 
\Xi_{hg}(\sigma) = \int_{-\infty}^{\infty} \xi_{hg}((x^2 +
\sigma^2)^{1/2}) \ dx 
\end{equation} 
We estimate $\Xi_{hg}$  using a method similar to
that described by Saunders et al (1992). For each HII galaxy  we
count the excess neighbours compared to a random distribution 
using \begin{equation} 
\label{eq:xihg} 
X_{hg}(\sigma) = {{ N_{hg}(\sigma)} \over {N_{hr}(\sigma) }}
-1 
\end{equation} 
This is related to $\Xi$ 
\begin{equation} 
\Xi(\sigma) \approx { {\int n(x) dx} \over {n(y)} } X(\sigma) 
\end{equation} 
where $n(x)$ is the average number of field galaxies per
steradian per Mpc along the line of sight 
in the background catalogue at distance $x$, and the
distance to the central HII galaxy is $y$.
The approximation is essentially the small-angle approximation, but
also involves several subtleties, as discussed by Saunders et al
(1992).
Note that the different distance to each HII galaxy means that
relation between $\sigma$ and $\theta$ is different for each centre,
and also the $ 1/n(y) $ leads to a different weighting of the pair
count from each centre.  This means that $\Xi_{hg}$ is not simply a
rescaling of $w_{hg}$.

If $\xi_{hg}$ is a power law in $r$, $\ \xi_{hg}=(r/r_0)^{-\gamma}$, then the
projected correlation function is given by 
$\ \Xi_{hg}(\sigma) = (\sigma/\sigma_0) ^{1-\gamma} \ \ $ where 
$\ \sigma_0 ^{\gamma-1} = r_0 ^{\gamma} \Gamma(\frac{1}{2} ) 
\Gamma( \frac{\gamma-1}{2} ) / \Gamma ( \frac{\gamma}{2} ) .\ $ 
Hence 
\begin{eqnarray}
\Xi_{hg}(\sigma)/\sigma & = 
\xi_{hg}(\sigma) \Gamma(\frac{1}{2} )  \Gamma(\frac{\gamma-1}{2})
/ \Gamma( \frac{\gamma}{2}) .
\label{eq:Xi} 
\end{eqnarray}

Our measurement of $\Xi_{hg}(\sigma)/\sigma$ is shown by the points 
in Figure~\ref{wsigmar}.  It is positive for $\sigma \simlt 1$Mpc,
showing that HII galaxies have more neighbouring galaxies than a
uniform distribution: they are not isolated systems.
As discussed in Section~3.1, we expect any sample of galaxies to have
more neighbours than a random distribution, so we have calculated the
expected $\Xi/\sigma$ assuming that HII galaxies are clustered in the
same way as normal galaxies.
Our prediction is given by equation~\ref{eq:Xi} with $\xi_{hg}(r) =
(r/5.1h^{-1})^{-1.71} $, which is a good approximation to the APM 
correlation function on small scales (Maddox \etal 1996). 
We have also subtracted a constant from the power-law to allow for the
integral constraint which applies to the data points.  
This is shown as the solid line in Figure~\ref{wsigmar}.
The observed values are consistent, although slightly lower than the
predictions. 
This slight discrepancy can be interpreted as a reflection of the fact
that the amplitude of the autocorrelation function for HII galaxies is
lower than for normal galaxies. Iovino Melnik and Shaver (1988) find
$r_0 = 2.7 h^{-1}$Mpc for the HII galaxy sample used here; the
corresponding $\Xi_{hh}$ is shown as the 
dashed  line in Figure~\ref{wsigmar}. 
If HII galaxies and normal galaxies follow the same underlying mass
distribution with differing bias levels, the cross-correlation
function should be the geometric mean of the two autocorrelation
functions, so $\xi_{hg} = \sqrt{\xi_{hh}\xi_{gg}}$.
The dotted line in Figure~\ref{wsigmar} shows the equivalent  $\Xi_{hg}$.

The lower amplitude emphasises the point made in Section 3.1, that
there is not a large excess of near neighbours around the HII
galaxies compared to normal galaxies.
Again this is incompatible with suggestion that HII galaxies are
triggered by tidal interactions with nearby low-mass galaxies.

\section{conclusions}\label{conclusion}


Our main results  are: 

\begin{enumerate}

\item Both the angular and projected correlation functions are
significantly positive, so HII galaxies are significantly clustered.
This is what you expect to find for any sample of galaxies. 

\item The angular measurements have large
uncertainties, but are consistent with the predictions
expected for a sample of normally clustered galaxies. 

\item The  projected measurements are marginally lower than 
the predictions expected for a sample of normally clustered galaxies,
and lie between the autocorrelation functions of normal galaxies and 
HII galaxies.


\end{enumerate}

Telles and Terlevich (1995) found that the space density of bright
galaxies within 1Mpc$^3$ of HII galaxies is a factor $\sim 4$ times
higher than expected for a random distribution, but $\sim 2 $ times
less than for a control sample of sample of Sc galaxies.
These results showed that HII galaxies are more clustered than a random
distribution, but slightly less clustered than normal galaxies.
The present work extends this analysis to much fainter apparent
magnitudes, measuring the density of galaxies to $M_{b_J} \sim -14.5$,
which correspond to very low mass galaxies and HI
clouds ($\sim 10^8$\msun).
Our results are in agreement with the  earlier studies, with the
additional conclusion  that HII galaxies do not have preferentially
faint, low-mass neighbours. 

We conclude that star-formation in these galaxies is not triggered by
tidal interactions and must have a different origin, possibly
associated with the formation and evolution of Super Stellar Clusters
in starbursts, as observed in the UV (Vacca 1994; Meurer \etal 1995;
Ho 1997) and in the Near-IR (Telles \etal 1999).

  
\normalsize
  

\section*{References}  

\def\ref{\noindent \hangindent 10pt \hangafter 1 }

\ref Balzano, V.A., 1983, \apj, 268, 602

\ref Brinks,E., 1990, in \,``Dynamics and Interactions of Galaxies\,'',  
ed. R. Wielen, Springer-Verlag, Heidelberg, p. 146  
  
\ref Campos-Aguilar,A., Moles,M., 1991, \aanda, 241, 358  
  
\ref Campos-Aguilar,A., Moles,M., Masegosa,J., 1993, \aj, 106, 1784  
  
\ref Hamilton, A.J.S., 1993, ApJ, 417, 19

\ref Hoffman, L., 1999, in "Dwarf Galaxies and Cosmology"
 eds. T. X. Thu\^an, C. Balkowski, V. Cayatte \& J. Tr\^an Thanh
 V\^an, Editions Fronti\`eres (Gyf-sur-Yvette, France) 

\ref Ho, L. 1997, \rmx, 6, 5

\ref Iovino, A, Melinick J. \& Shaver P., 1988, \apjl, 330, L17

\ref Kaiser, N., 1986, \mn, 219, 785

\ref Loveday, J., Tresse, L. \& Maddox, S., 1999, in preparation

\ref MacAlpine G.M. \&  Williams G.A., 1981, \apjs, 45, 113

\ref Maddox, S.J., Sutherland, W.J. Efstathiou, G., and Loveday, J.,
1990, \mn, 243, 692

\ref Maddox, S.J., Efstathiou, G. and Sutherland, W.J., 1996, \mn, 283 1227.

\ref Markarian, B.E., Lipovetskii, V.A. \& Stepanyan, Dzh.A., 1981,
Astrofisika, 17, 619

\ref Melnick,J., 1987, in {\it \,``Starburst and Galaxy Evolution\,''}, 
eds. T.X.Thuan, T.Montmerle \& J.Tran Thanh Van, editions Fronti\`eres Gif 
Sur Yvette, France, p. 215

\ref Melnick,J.,Mirabel,I.F., 1990, \aanda, 231, L19  

\ref Meurer, G.R., Heckman, T.M., Leitherer, C., Kinney, A., Robert, C.,
Garnett, D.R., 1995, \aj. 110, 2665. 

\ref Peebles P.J.E., 1980, {\it The Large-Scale Structure of the
Universe}, Princeton University Press, Princeton.

\ref Pustil'nik,S.A., Ugryumov,A.V., Lipovetsky,V.A., Thuan,T.X., Guseva, N.,
1995. \apj, 443, 499. 

\ref Rao, S. \& Briggs, F., 1993, \apj, 419, 515
  
\ref Rosenberg,J.L., Salzer,J.J., Moody,J.W., 1994, \aj, 108, 1557

\ref Saunders, W., Rowan-Robinson, M. and Lawrence, A., 1992, \mn, 258, 134

\ref Smith, M.G., Aguirre, C.  \& Zemelman, M., 1976, \apjs, 32, 217

\ref Smith R.J., Boyle B.J., and  Maddox S.J., 1995, \mn. 277, 270

\ref Soifer, B. T., Sanders, D. B., Madore, B. F., Neugebauer, G.,
Danielson, G. E., Elias, J. H., Lonsdale, Carol J., Rice, W. L., 1987,
\apj, 320, 238

\ref Suntzeff, N.B., Phillips, M.M., Covarrubias, R., Navarrete, M.,
Peres, J.J.P., Guerra, A., Acevedo, M.T., Doyle, L.R., Harrison, T.,
Kane, S., Long, K.S., Maza, J., Miller, S., Piatti, A.E., Claria,
J.J., Ahumada, A.V., Pritzl, B., Winkler, P.F., 1998, astro-ph/9811205

\ref Taylor,C.L.,Brinks,E., Grashuis, R.M. \& Skillman,E.D., 1995, \apjs, 99, 427

\ref Taylor,C.L.,Brinks,E., Grashuis, R.M. \& Skillman,E.D., 1996, \apjs, 102, 189 (erratum)

\ref Taylor,C.L 1997, \apj, 480, 524

\ref Telles,E. \& Terlevich,R., 1995, \mn, 275, 1
 
\ref Telles,E. \& Terlevich,R., 1997, \mn, 286, 183
 
\ref Telles,E., Melnick,J. \& Terlevich,R., 1997, \mn, 288, 78

\ref Telles,E., Tapia,M., Terlevich,R., Kunth,D. \& Sampson,L., 1999,
 in: K.A. van der Hucht, G. Koenigsberger \& P.R.J. Eenens (eds.),
 ``Wolf-Rayet Phenomena in Massive Stars and Starburst Galaxies'',
 Proc. IAU Symposium No. 193 (San Francisco: ASP), in press

\ref Terlevich, R., Melnick, J., Masegosa, J., Moles, M. \& Copetti, M.V.F.,  
    1991, \aandas, 91, 285 (SCHG)  

\ref Vacca, W.D., 1994, in ``Violent Star Formation'', ed Tenorio-Tagle, p 297

\ref Vilchez,J.M., 1995, \aj, 110, 1090

\ref Zwaan, M., 1999, in "Dwarf Galaxies and Cosmology" proceedings
of the XVIIIth Moriond astrophysics meeting, 
eds. T. X. Thu\^an, C. Balkowski, V. Cayatte \& J. Tr\^an Thanh V\^an,
Editions Fronti\`eres (Gyf-sur-Yvette, France) 

\ref Zwaan, M.A., Briggs, F.H., Sprayberry, D., Sorar, E., 1997, \apj, 490, 173
 
\ref Zwicky, F., 1971, Catalogue of Selected Compact Galaxies
 and Post-Eruptive  Galaxies, published by the author, Switzerland

  
\end{document}